\documentclass[referee]{raa}            

\usepackage{graphicx,times}             
\usepackage{float}                                 

\begin{document}

   \title{LAMOST CCD camera-control system based on RTS2
}

   \volnopage{Vol.0 (200x) No.0, 000--000}      
   \setcounter{page}{1}          

   \author{Yuan Tian
      \inst{1,3}
   \and Zheng Wang
      \inst{1,3}
   \and Jian Li
      \inst{1}
   \and Zi-Huang Cao
      \inst{1,3}
   \and Wei Dai
      \inst{2}
   \and Shou-Lin Wei
      \inst{2}  
   \and Yong-Heng Zhao
      \inst{1,3}
   }

   \institute{Key Laboratory of Optical Astronomy, National Astronomical Observatories, Chinese Academy of Sciences,
             Beijing 100101, China; {\it tianyuan@bao.ac.cn}\\
        \and
             Computer Technology Application Key Lab of Yunnan Province, Kunming University of Science and Technology, Kunming 650500, China
        \and
             University of Chinese Academy of Sciences, Beijing 100049, China
   }

   \date{Received~~2009 month day; accepted~~2009~~month day}

\abstract{The Large Sky Area Multi-Object Fiber Spectroscopic Telescope (LAMOST) is the largest existing spectroscopic survey telescope, having 32 scientific charge-coupled-device (CCD) cameras for receiving spectra. Stability and automation of the camera-control software are essential, but cannot be provided by the existing system. The Remote Telescope System 2nd Version (RTS2) is an open-source and automatic observatory-control system. However, all previous RTS2 applications have concerned small telescopes. This paper focuses on implementation of an RTS2-based camera-control system for the 32 CCDs of LAMOST. A virtual camera module inherited from the RTS2 camera module is built as a device component working on the RTS2 framework. To improve the controllability and robustness, a virtualized layer is designed using the master-slave software paradigm, and the virtual camera module is mapped to the 32 real cameras of LAMOST. The new system is deployed in the actual environment and experimentally tested. Finally, multiple observations are conducted using this new RTS2-framework-based control system. The new camera-control system is found to satisfy the requirements for LAMOST automatic camera control. This is the first time that RTS2 has been applied to a large telescope, and provides a referential solution for full RTS2 introduction to the LAMOST observatory control system.
}

   \authorrunning{Y.Tian  et al.}            
   \titlerunning{LAMOST CCD camera-control system based on RTS2}  

   \maketitle

%
%
\section{Introduction}           
\label{sect:intro}

The Large Sky Area Multi-Object Fiber Spectroscopic Telescope (LAMOST, also named the “Guo-Shou-Jing Telescope”) is a meridian reflecting Schmidt (Wang-Su-type,~\cite{cui2012}) telescope. LAMOST, as a Chinese national scientific research facility, is operated by the National Astronomical Observatories, Chinese Academy of Sciences (NAOC), and is located at Xinglong Station, which is a national facility open to the astronomical community. As the telescope having the highest spectrum acquisition rate in the world, LAMOST has already successfully observed and processed more than 7.68 million spectra \footnote{http://dr4.lamost.org}, and is a powerful tool for wide-field and large-sample astronomy (~\cite{zhao2015}).

The main spectroscopic survey instruments of LAMOST are 16 low-resolution spectrographs (~\cite{hou2010}), each of which contains two scientific charge-coupled-device (CCD) cameras (English Electric Valve Company (EEV) 203-82, 4K x 4K pixels); thus, 32 CCDs are used for observation (~\cite{zou2006}). In 2009, the present authors developed CCD control software to coordinate procedures and to collect status information for the 32 CCDs (~\cite{deng2010}, ~\cite{deng2011}). A schematic diagram of the spectrographs and CCDs is shown in Figure 1.

\begin{figure}[!htbp]
\centering
\includegraphics[width=\textwidth, angle=0]{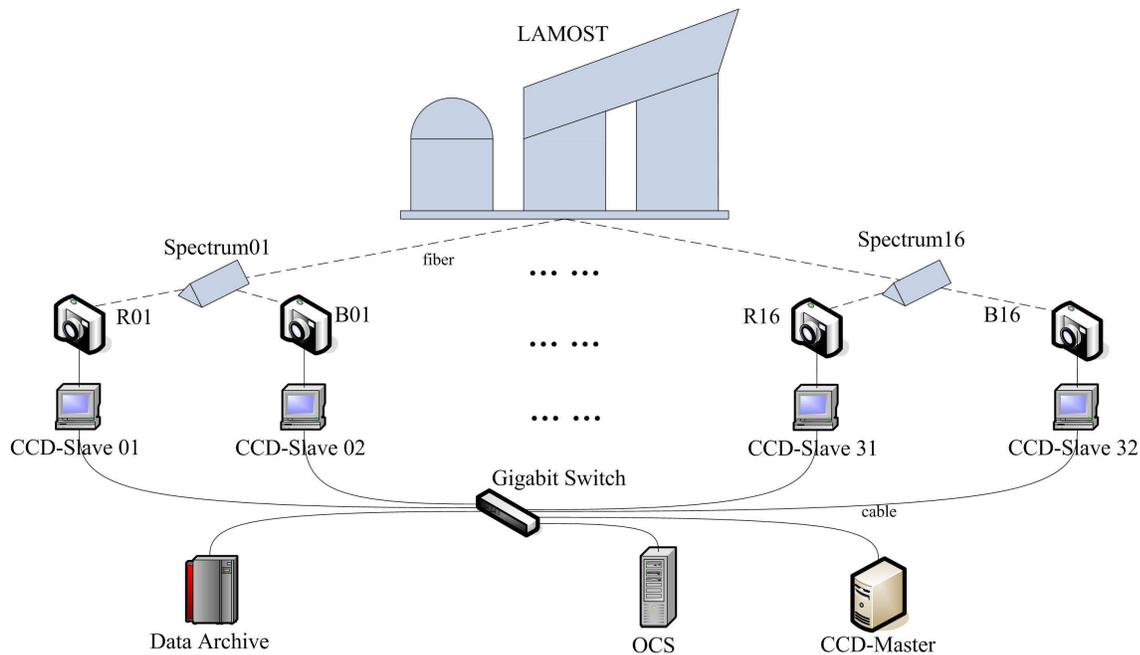}
\caption{spectrographs and CCDs in LAMOST}
\label{Fig1}
\end{figure}

The developed software has worked for 8 years, and satisfied its basic design goal. However, this software is excessively complex, maintenance is tedious, and full integration into the LAMOST observatory control system (OCS) is difficult. With the development of the LAMOST spectrum survey, in order to improve software stability and observation efficiency, automation and robustness have become new requirements. It is apparent that the existing CCD control software cannot meet these challenges; thus, a software upgrade is required.

Computer-controlled telescope technology began to appear in the 1980s, following the introduction of personal computers. The considerable advancements in modern information technology since the beginning of the 21st century have contributed to its rapid development, and the goals of automation and robustness have now been achieved for a large number of telescopes. The development history of computer-controlled telescope technology can be divided into four stages: Automated Scheduled Telescopes, Remotely Operated Telescopes, Robotic Autonomous Observatories, and Future Robotic Intelligent Observatories (~\cite{castro2010}). Thus, the present study targets development related to the third stage, Robotic Autonomous Observatories.

Following an in-depth study of telescope automation control technology and software (~\cite{cui2013}), we chose Remote Telescope System 2nd Version (RTS2), a representative third-generation control technology, for introduction to the LAMOST OCS. The first stage of this upgrade project involves implementation of RTS2 to drive the 32 CCDs, as one of the core OCS components. RTS2 is an open-source package for remote telescope control under the Linux operating system and is written in C++. It is designed to run in fully autonomous mode, which can be used to coordinate all autonomous scheduling, telescope pointing, data acquisition, instrument hardware monitoring, etc. (~\cite{peter2006}).

The autonomous capabilities and robustness of RTS2 benefit from its object-oriented design and reasonable class hierarchy. With regard to communication, the code is divided into three levels: the communication code, device-type specific library, and native drivers. The communication code is based on transmission control protocol/internet protocol (TCP/IP), using custom protocol and heartbeat technology, and can handle various network transport transactions and error recovery. The device-type library, which summarizes the common steps of the general telescope equipment, handles device commands (initiation of mount movement and exposure, etc.). The native drivers can drive various device types (CCDs, mounts, sensors, etc.). By inheriting the base device-type class and overriding parts of the virtual function interfaces, a software re-developer can quickly create a new dedicated device driver, which can then be customized for application to various tasks.

As a result of continuous improvement, RTS2 has evolved into a modular package supporting a variety of hardware devices and providing reliable general-purpose observatory control software. Thus, RTS2 is currently running on various small telescope setups and laboratory equipment control systems worldwide\footnote{http://rts2.org/wiki/obs:start} (~\cite{peter2012-1}, ~\cite{zhang2016}). However, because of its complexity, RTS2 has not been successfully applied to large-aperture telescopes to date.

The aim of this study is to upgrade the existing LAMOST camera-control software to adapt to the RTS2 framework, as the first stage of the ultimate project aim of introducing RTS2 to the LAMOST OCS. Here, the developed RTS2-based camera-control system is described in detail and implemented in the actual environment, with multiple test observations being conducted..


\section{Software Design Architecture and Implementation}
\label{sect:SDAI}
As the 32 CCDs are the most important acquisition components of the LAMOST spectral equipment, their control system must be stable, automated, and easy to maintain. To satisfy these requirements, we have introduced an RTS2-based software framework to the LAMOST CCD control system to upgrade the existing software. The CCD control system is shown in the context of the overall LAMOST software framework in Figure 2.

\begin{figure}[!htbp]
\centering
\includegraphics[width=\textwidth, angle=0]{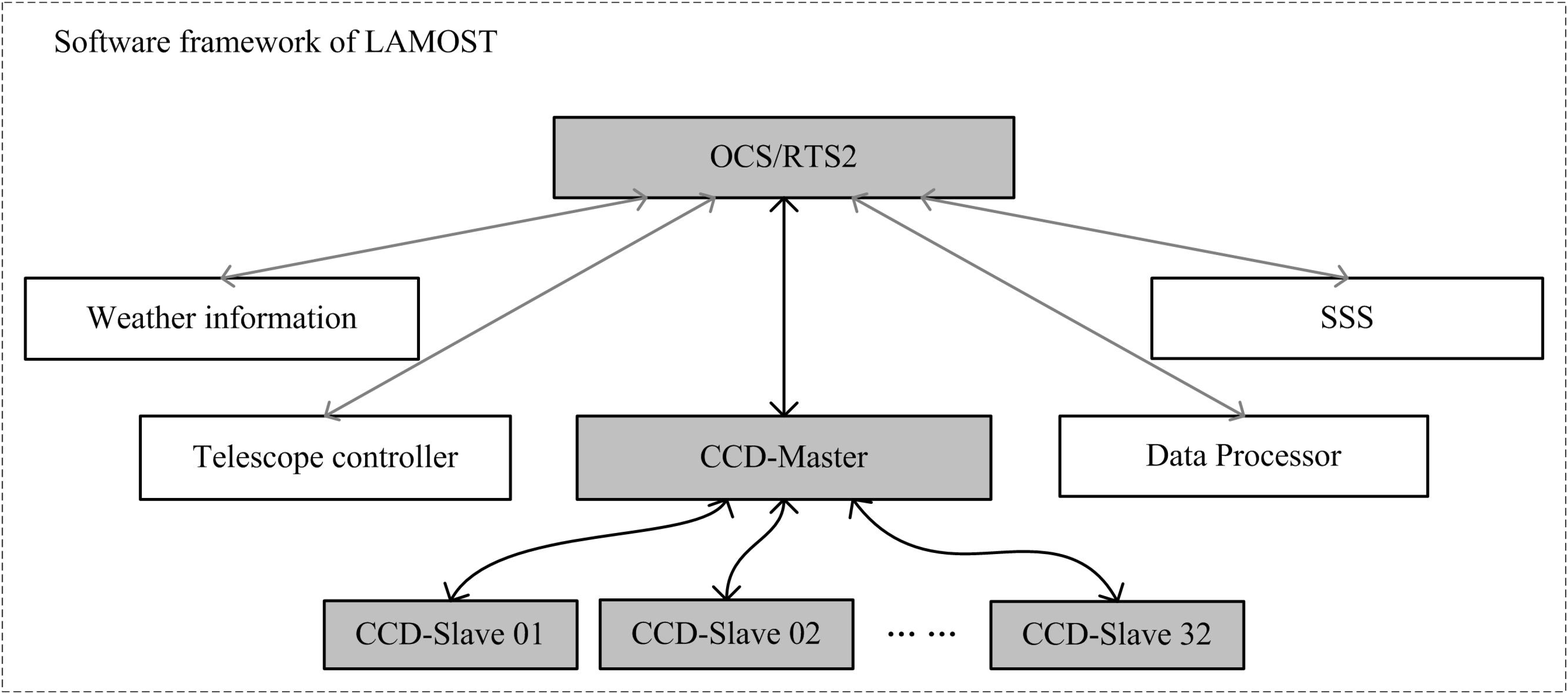}
\caption{spectrographs and CCDs in LAMOST (SSS: Survey Strategy System, which is responsible for generating observation plans and arranging observation sequences)}
\label{Fig2}
\end{figure}

An object-oriented design method is employed for RTS2, which can support many kinds of cameras. A method for application of RTS2 for customized devices also exists\footnote{http://rts2.org/wiki/doku.php?id=code:camera\_driver}, and this technology is very simple and convenient to use. In contrast, the existing LAMOST CCD control application is more complex. Note that the type of CCD camera employed by LAMOST is not supported by RTS2 directly. Further, no use case involving the control and coordination of such a large number of cameras by RTS2 has been reported to date. Therefore, it was necessary to design the system meticulously and to perform careful testing.

\subsection{Design Architecture}
In order to bridge RTS2 with the 32 EEV CCDs (real cameras) of LAMOST, the designed software is divided into three layers: the center controller layer (OCS, integrated with RTS2), virtual device layer (named “CCD-Master”), and real control layer (“CCD-Slave” and Universal Camera, also named UCAM). UCAM is a stand-alone driver software for scientific CCD camera develeped by the LICK Observatory.  Becase of  its good versatility and stability, LAMOST used it as the driver for EEV CCDs (~\cite{zou2006}, ~\cite{jia2010}). The software architecture is depicted in Figure 3.

\begin{figure}[!htbp]
\centering
\includegraphics[width=\textwidth, angle=0]{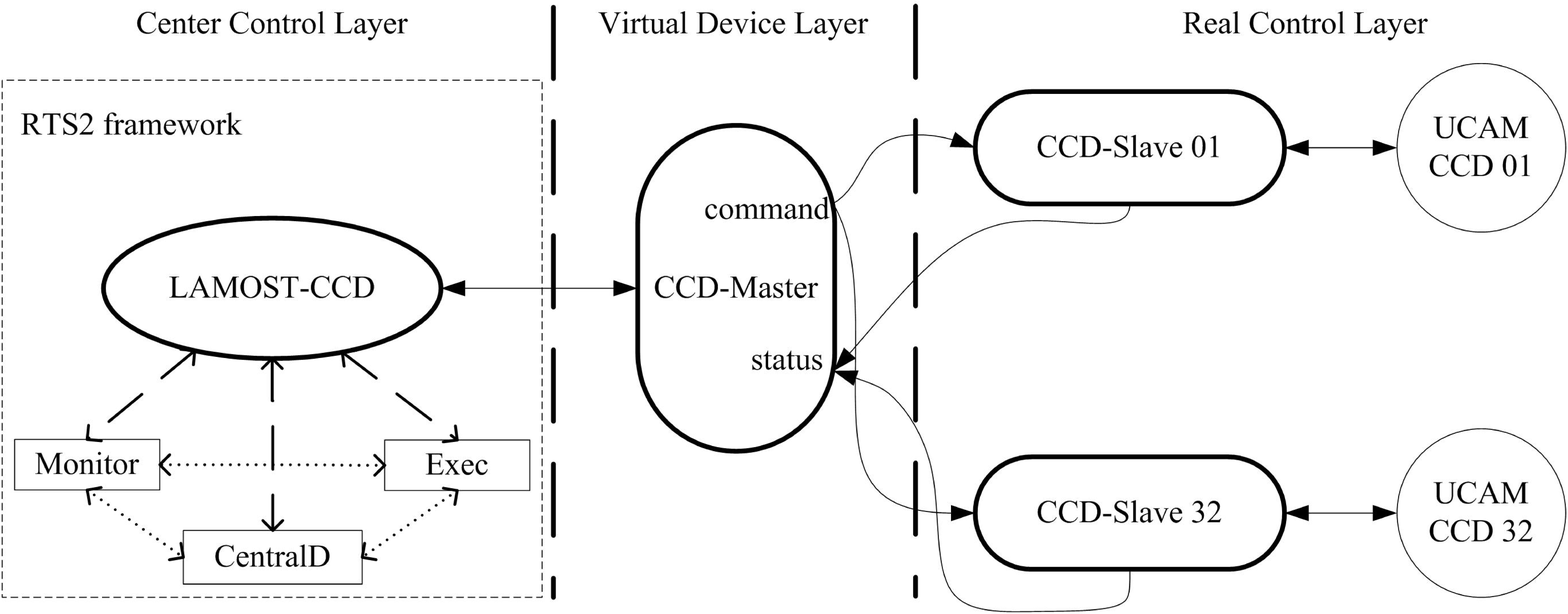}
\caption{Overall software architecture (CentralD: Name resolver and observatory housekeeper; EXEC: “Executor” device scheduler)}
\label{Fig3}
\end{figure}

In the center control layer, a customized RTS2 camera component (named “LAMOST-CCD”) is created, which processes related RTS2 operations. The master-slave paradigm is used in the virtual device layer and real control layer. In the virtual device layer, a master, i.e., CCD-Master, is designed, which distributes commands to and collects status information from the slaves. It also maintains a connection with LAMOST-CCD, to receive command-and-response general status information. In the real control layer, CCD-Slave is implemented as a slave, which is a bridge connecting CCD-Master and the EEV CCD stand-alone driver software, i.e., UCAM.

This design scheme preserves the advantages of RTS2, e.g., target scheduling, control automation, and fault tolerance. We also attempted to retain the existing software interface. Separation of the command/status stream was used to achieve a simple design, accelerating the code development progress.

\subsection{Implementation of customized camera class in RTS2 (LAMOST-CCD)}
A new RTS2 camera class was created to control the LAMOST CCDs, named “LAMOST-CCD.” This is an RTS2 camera module and inherits from rts2camd::Camera. The inheritance diagram of our camera class in the RTS2 framework is shown in Figure 4.

\begin{figure}[!htbp]
\centering
\includegraphics[width=\textwidth, angle=0]{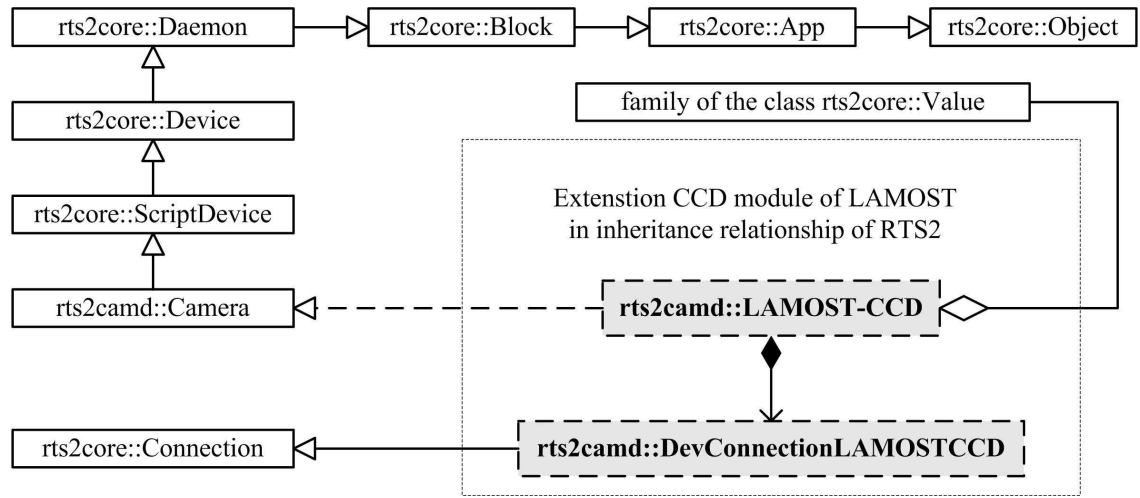}
\caption{Inheritance diagram of customized camera module class in RTS2 framework}
\label{Fig4}
\end{figure}f

The rts2camd::Camera class is a generic camera module class provided by RTS2, being an abstract class for all kinds of cameras. It simply summarizes the general control steps and statuses of astronomy cameras and reserves a large number of (virtual functions) interfaces (~\cite{wei2014}). In this project, LAMOST-CCD overrides a small number of virtual functions. These virtual functions can transmit special commands and statuses defined by the authors (for details of these custom commands, see Section 3).

The RTS2 framework implements connections between its modules using TCP sockets, and uses its own communication protocol. The core communication class is rts2core::Connection (~\cite{peter2008}). To satisfy the communication requirement of RTS2, we created a connection class named “DevConnectionLAMOSTCCD,” which inherits from rts2core::Connection. This class is used to manage the connection between LAMOST-CCD and CCD-Master. We override its interfaces using functions such as processLine() and setState(). The detailed unified modeling language (UML) diagram is shown in Figure 5.

\begin{figure}[!htbp]
\centering
\includegraphics[width=\textwidth, angle=0]{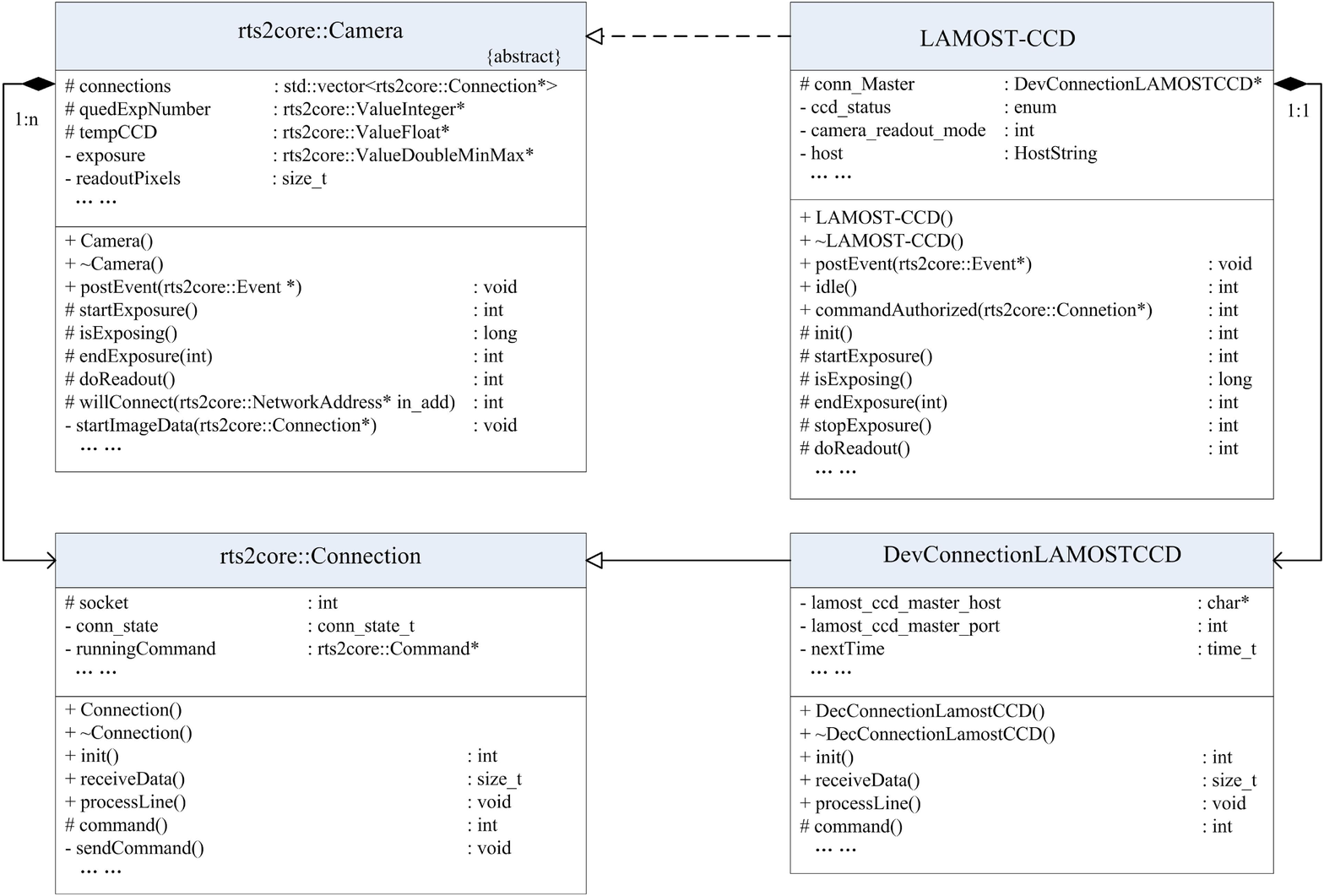}
\caption{LAMOST-CCD class UML diagram}
\label{Fig5}
\end{figure}

Following the RTS2 rules, when LAMOST-CCD runs on a separate process, it attempts to register itself into the RTS2 CentralD (the name resolver and observatory housekeeper) on rts2core::Connection. Meanwhile, it creates a DevConnectionLAMOSTCCD object to connect to CCD-Master. When this connection has been established, LAMOST-CCD adds this DevConnectionLAMOSTCCD object to its connections list. Subsequently, LAMOST-CCD functions in the same manner as any other RTS2 device module.

When the RTS2 executor or clients send a command to LAMOST-CCD, the latter transmits a command to CCD-Master through the TCP connection managed by DevConnectionLAMOSTCCD. When CCD-Master responds with aggregate statuses, DevConnectionLAMOSTCCD translates these statuses into custom events (EVENT\_LAMOST\_EXPOSURE\_START, EVENT\_LAMOST\_READOUT\_END, etc.), and delivers these events to LAMOST-CCD. Finally, LAMOST-CCD, as a RTS2 device module, changes its own state and performs the appropriate post-processing.

As RTS2 provides a Daemon-running framework, connection management framework, event handling mechanism, we can implement the LAMOST-CCD class very simply and rapidly.

\subsection{CCD-Master}
In order to map the 32 LAMOST CCDs to the RTS2 camera module, we added a virtual layer between them and implemented a master-slave program paradigm. 

CCD-Master has two tasks. As the first task, CCD-Master (as a socket server) accepts connections and receives commands from LAMOST-CCD. Then, it processes these commands (translates, sets special parameters, etc.). Finally, it connects to the 32 CCD-Slaves (as a socket client) and distributes these commands. As the second task, CCD-Master (as a socket server) accepts connections and receives status messages from the CCD-Slaves. Then, it judges whether all cameras are working well. Finally, it provides a total status report to LAMOST-CCD as one virtual camera. The internal structure of CCD-Master is shown in Figure 6. As a simple synchronous mode through socket communication is used to ensure command and status transmissions are not blocked by each other, CCD-Master implements two processes separately, one for command and the other for status.

\begin{figure}[!htbp]
\centering
\includegraphics[width=\textwidth, angle=0]{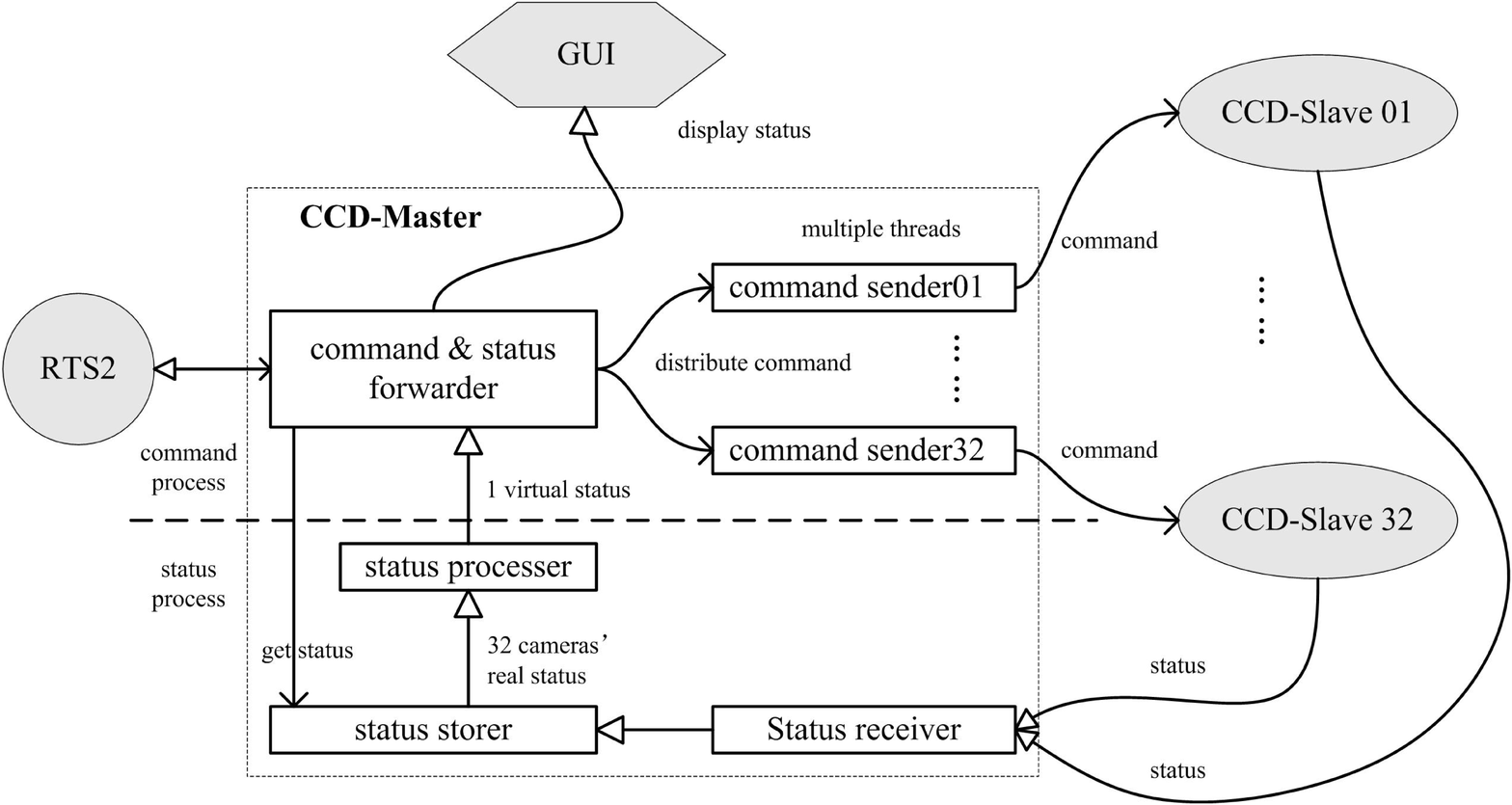}
\caption{CCD-Master internal structure diagram (GUI: Graphical user interface)}
\label{Fig6}
\end{figure}

In an ideal scenario, CCD-Master should accurately distribute commands to the 32 cameras simultaneously, especially when the commands are “start/stop exposure.” However, because of the program operation principles, this cannot be achieved very precisely. Fortunately, the nocturnal observations conducted at LAMOST typically involve long-term exposure (600 s $-$ 1800 s), and the UCAM time sensitivity is 0.01 s. Therefore, a TCP-based multi-thread design plan for command distribution was adopted. This plan is stable and low-cost, and satisfies the engineering requirements well. The time difference between multi-threads can be negligible in this case. Note that this approach was examined in experiment in this study, and detailed results are presented in Section 4.1.

In the status process, one thread is used, because the status-collection time accuracy is not critical. We added a storer to cache the status results for the 32 cameras. When RTS2 send command ``Get Status'', the CCD-Master's command \& status forwarder (in the command process) transfer this command to the storer. Then the storer sends the status results for the 32 cameras back to the status processer, and the status processer generates a virtual status. Finally, the status sender obtains this virtual status and returns it to LAMOST-CCD (RTS2). The status reception performance of CCD-Master is reported in Section 4.2.

A graphical user interface (GUI) was implemented for CCD-Master, which provides observers with the option to obtain more detailed information on the real cameras. The GUI is shown in Section 4.3.

\subsection{CCD-Slave}
The internal structure of CCD-Slave is shown in Figure 7. This design can retain the original interfaces of the existing software. This concept is very important when upgrading the existing software of a large telescope that has been running stably for a long time. CCD-Slave is a bridge between CCD-Master and UCAM, which is located in the personal computer (PC) of each CCD controller.

\begin{figure}[!htbp]
\centering
\includegraphics[width=\textwidth, angle=0]{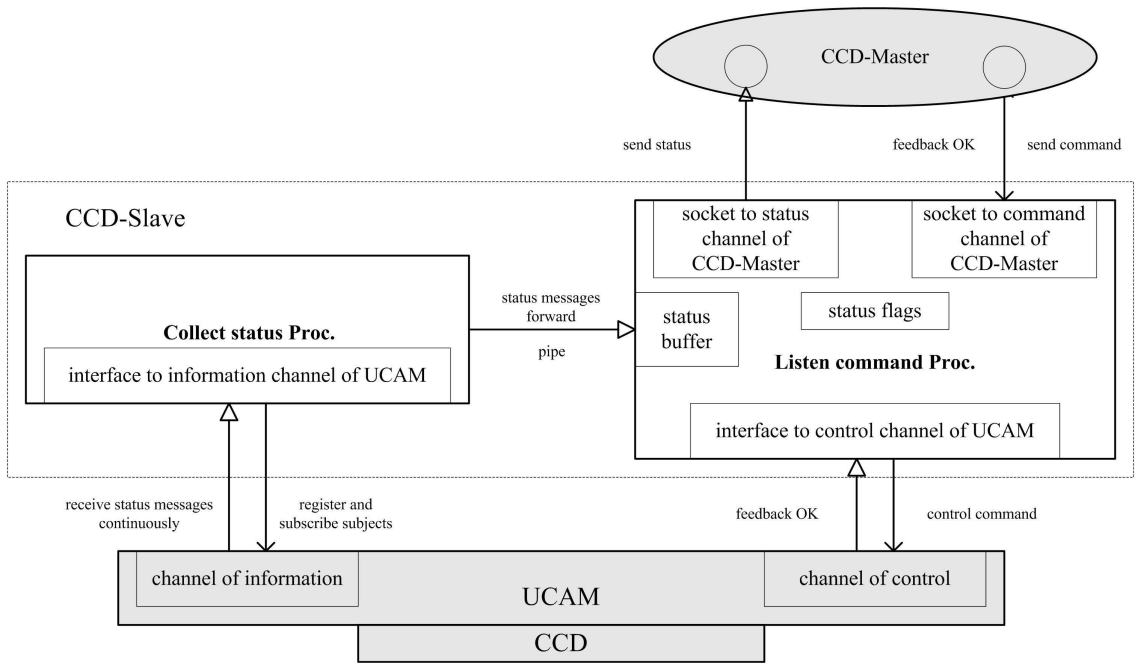}
\caption{CCD-Slave internal structure diagram}
\label{Fig7}
\end{figure}

Like CCD-Master, CCD-Slave also has two separate processes, one of which manages commands, while the other handles status messages. The command listener creates a socket and receives commands from CCD-Master, before translating the commands to a special string (or binary codes) that can be understood by UCAM. When CCD-Master sends a "Get Status" command, the command-listener status flag changes to TRUE. Then, the command listener connects to the CCD-Master status endpoint and sends the status information, which is cached in "status buffer."

The status collector process registers itself into the UCAM information channel and subscribes to certain status topics; then, it receives messages continuously. When a new status message is received, the status collector pushes it into the status buffer through the pipe, which is a bridge that connects the two processes. The pipe is created when CCD-Slave is initiated.

\section{Internal Communication Protocol}
\label{sect:protocol}
RTS2 employs its own communication protocol (~\cite{peter2008}), which is based on sending American Standard Code for Information Interchange (ASCII) strings over TCP/IP sockets. It is fast, simple, and robust. As a device module, LAMOST-CCD uses the existing RTS2 protocol to communicate with other RTS2 modules. To communicate with CCD-Master, LAMOST-CCD uses customized commands and statuses based on the RTS2 protocol. Table 1 lists various custom-defined commands and status messages.

For convenience, we used a key-value pattern to transport and store various parameters. Implementation of this approach is very simple and this method is also very flexible. Using this technique, we could easily add and modify parameters throughout the entire code development process. For example, when LAMOST-CCD receives a "set parameters" command from the RTS2 internal module, the former then translates the command according to Table 1 and adds a "key-value" pair (e.g., $<$selected=all$>$). Then, LAMOST-CCD sends the translated command to CCD-Master. CCD-Master receives that command, parses it, obtains the value of "selected," and then forwards the command to all CCD-Slaves. Each CCD-Slave receives and parses the command and then drives its CCD camera.

\begin{table}
\begin{center}
\caption[]{ Custom-defined commands and status messages based on RTS2 protocol}\label{Tab:custom-protocol}


 \begin{tabular}{clcl}
  \hline\noalign{\smallskip}
Format &  Explanation & Sender & Receiver                    \\
  \hline\noalign{\smallskip}
GV $<$key$>$                                  & Get parameters              & LAMOST-CCD  & CCD-Master\\
XV $<$key$>$  $<$value$>$	  & Set parameters               & LAMOST-CCD  & CCD-Master\\
EX                                                        & Exposure start                  & LAMOST-CCD  & CCD-Master\\
SX                                                        & Exposure stop                  & LAMOST-CCD  & CCD-Master\\
RD                                                       & Readout                             & LAMOST-CCD  & CCD-Master\\
A   $<$Message$>$                      & Warning information        & CCD-Master       & LAMOST-CCD\\
+   $<$Message$>$                      & Success information        & CCD-Master       & LAMOST-CCD\\
-   $<$Message$>$                      & Failed information             & CCD-Master       & LAMOST-CCD\\
$\cdots\cdots$                              & Other control command  & LAMOST-CCD   & CCD-Master\\
  \noalign{\smallskip}\hline
\end{tabular}
\end{center}
\end{table}

\section{Software Deployment and Testing}
\label{sect:testing}
After the software development was completed, the software was deployed in the real camera control environment of LAMOST (shown in Figure 1). The 32 LAMOST cameras (EEV CCDs) obtain light signals from 4000 optical fibers. Each camera has a controller PC, which has an Intel(R) Core(TM) i3-2100 3.10-GHz central processing unit (CPU) and 4 GB of memory. The CCD-Master server has a 2-way Intel Xeon 2.53-GHz processor (containing 16 logical CPU cores in total) and a 12 GB memory. The hardware configuration of RTS2 LAMOST-CCD PC is identical to that of the camera-control PC. All computers are connected to a Gigabit Switch and a CentOS 6.8 operating system is installed on each computer.
We designed a series of experiments to test the performance of the new control software, to verify whether this software can satisfy the actual observation requirements of LAMOST. The results are reported in the following subsections.

\subsection{Command distribution performance}
We designed the first experiment to test the performance of the TCP-based command distribution, and to determine whether the new system satisfies the engineering requirements for controlling 32 real CCD cameras. Note that the performance of the existing User Datagram Protocol (UDP)-based system is also discussed in this subsection, for comparison with the upgraded system. We produced alphabetic character command data (8 B, 12 B, ..., 1 KB), and distributed them via both the existing software and the new system. To guarantee experimental accuracy, the simulated commands of every size were each tested for five loops, where each loop distributed the command 1 million times. The averaged results for the distribution speed and network bandwidth occupancy are shown in Figure 8 and Figure 9, for the existing software and new system, respectively.

\begin{figure}[!htbp]
\centering
\includegraphics[width=\textwidth, angle=0]{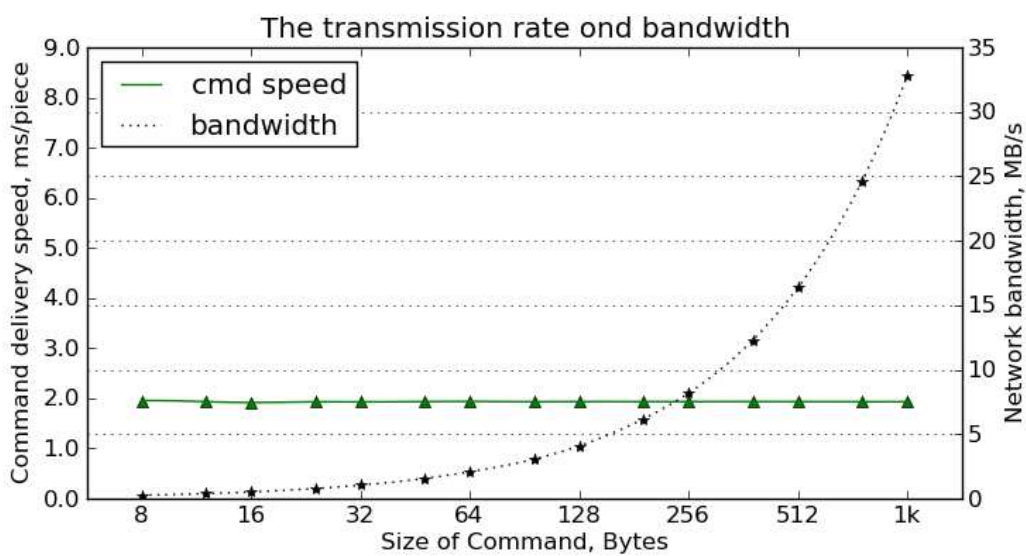}
\caption{ Command distribution performance of existing software, based on transmission rate and bandwidth as functions of command size}
\label{Fig8}
\end{figure}

\begin{figure}[!htbp]
\centering
\includegraphics[width=\textwidth, angle=0]{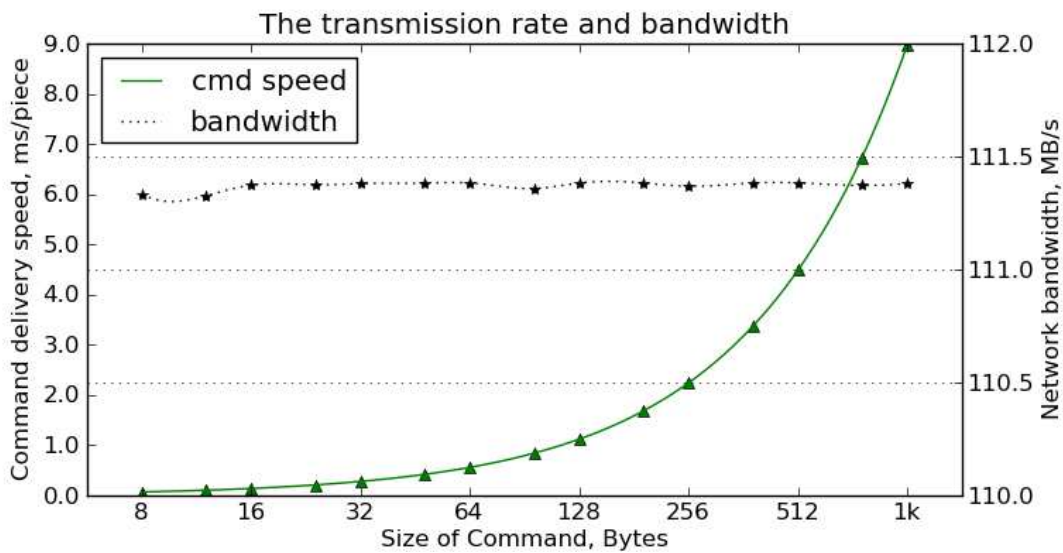}
\caption{Command distribution performance of new system, based on transmission rate and bandwidth as functions of command size}
\label{Fig9}
\end{figure}

In the existing UDP-based software, the command distribution speed is constant at 1.94 ms$/$piece, and the network occupancy increases with the command packet size. However, in the new, TCP-based system, because multi-thread technology is employed, the network occupancy is always close to the ultimate transmission speed (the gigabit switch ultimate speed is 1 billion bps $/$ 8 $\approx$ 119.21 MBps), and the command delivery speed increases with the command pack size. From comparison of the two figures, it is apparent that the new system exhibits superior performance to the existing system when the command pack size is less than 200 B. In contrast, when the command pack size exceeds 200 B, the new system exhibits inferior performance to the existing software. However, even when the command size is 1 KB, the delivery speed of the new system is as high as 8.94 ms$/$piece, which is far less than 100 ms , the UCAM time sensitivity. Therefore, the command distribution speeds of both software systems can satisfy the actual engineering requirements of LAMOST.
In real observation scenarios, our camera commands are usually smaller than 200 B; thus, use of TCP is preferable to UDP. Further, using TCP, there is no risk of packet loss or out-of-order packets. Moreover, TCP provides more options for the already mature library implementation, which is expected to reduce the maintenance and redevelopment workload significantly. Considering all the above, we can conclude that the new system can more effectively satisfy the actual engineering requirements of LAMOST.

\subsection{Status transmission performance}
In the second experiment, we tested the status collection performance of the new system to determine whether it satisfies the engineering requirements for collecting and processing the status information of the 32 real CCD cameras. Status data packets (8 B, 12 B, ..., 1 KB) were created for each of the 32 cameras. Then, all these status packets were sent to our new system simultaneously. As previously, to guarantee the experiment accuracy, the simulated status packets of each size were tested for five loops, and each loop involved one million distributions. The averaged results for the transmission speed and occupancy network bandwidth are shown in Figure 10.

\begin{figure}[!htbp]
\centering
\includegraphics[width=\textwidth, angle=0]{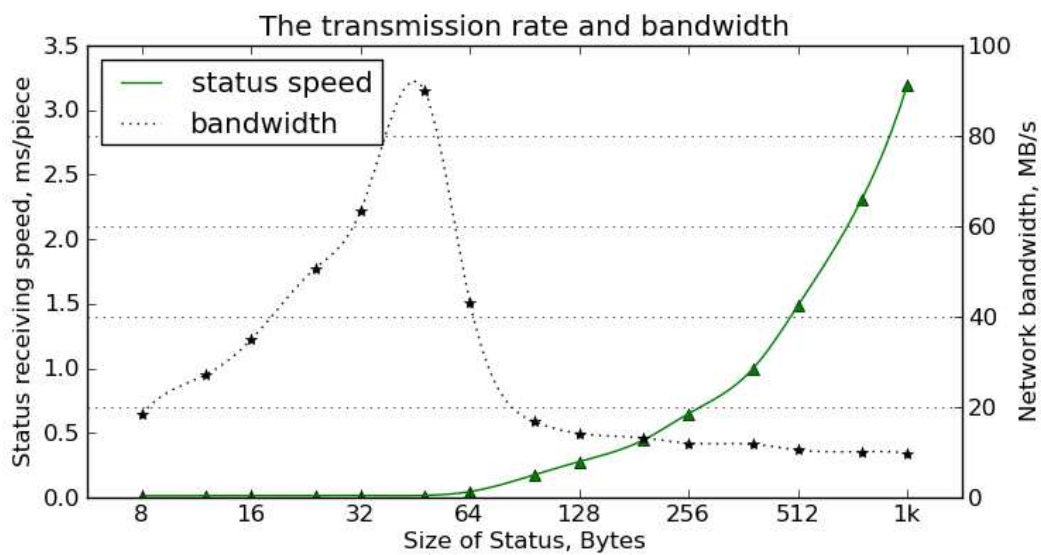}
\caption{Status collection performance of new system}
\label{Fig10}
\end{figure}

For simple development and maintenance, we designed a single thread and one endpoint only to receive and process all status packets. From Figure 10, the status processing speed increases with the status packet size, as for the command experiment. However, the network occupancy is not very high, and exhibits a peak at a status packet size close to 48 B. The reason for this phenomenon is that the software is limited by input/output (IO) blocking and sender judgment logic. However, even when the status packet size is 1 KB, the status receiving and processing speed is only 3.2 ms$/$piece. Because it is not critical for the status processing requirements to be satisfied in real observation scenarios, the single-camera status generation rate is approximately second level. Thus, the status processing performance of our developed system is adequate.

For details of the reception performance and packet loss rate of the existing system, please see ~\cite{dong2009}. As this information has already been published, it is not described again here.

\subsection{Automatic observation}
We wish to introduce RTS2 to the LAMOST OCS so as to implement its automatic telescope control capability and robustness, thereby overcoming the shortcomings of the existing camera-control software. Therefore, a third experiment was designed to test the automation and robustness of the new system.

\begin{figure}[H]
\centering
\includegraphics[width=\textwidth, angle=0]{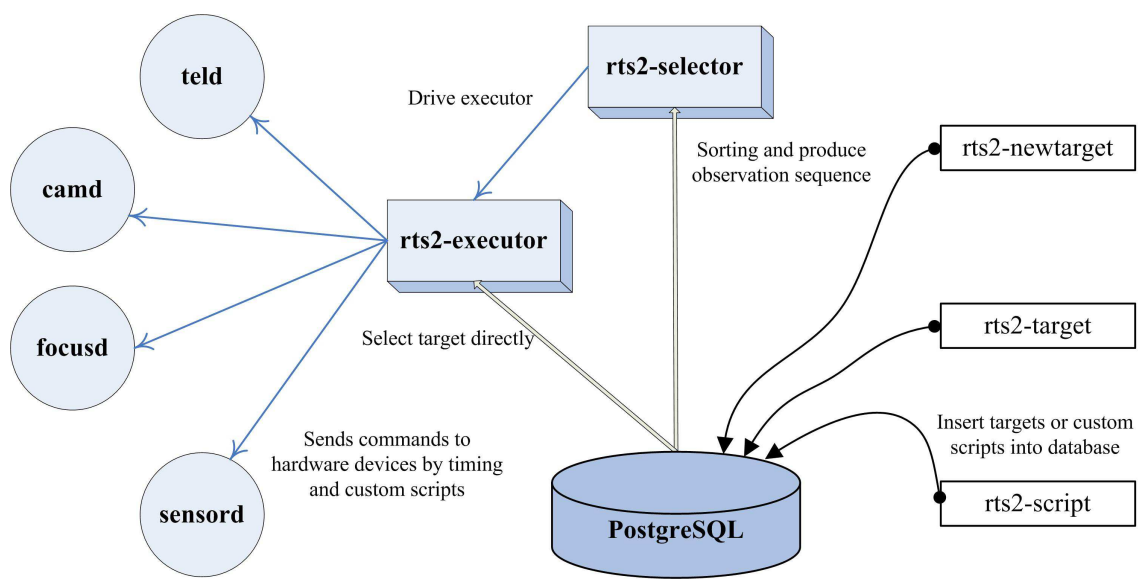}
\caption{The schematic diagram of RTS2 automatic observation}
\label{Fig11}
\end{figure}

RTS2 provides an observation decision-maker named "selector" (SEL), which can automatically select targets from a database according to various strategies (~\cite{peter2012-2}). And the targets in database can be insert or updated by RTS2scripts, e.g. rts2-newtarget, rts2-target. RTS2 also provides a device scheduler named "executor" (EXEC). These two core modules, along with the CentralD status synchronizer, guarantee the RTS2 automation capability. In addition, all RTS2 modules have pluggable implementation. Thus, the modules run independently and are connected dynamically. This property provides RTS2 with fault tolerance and robustness. It is possible to remove and add devices during observation without affecting the entire observation. Finally, RTS2 also provides a large number of dummy device modules for various types of astronomical equipment. The relationship between RTS2 modules for automatic observation is shown in Figure 11. Therefore, when our camera control software was realized, we could register our new system in the RTS2 environment and immediately begin simulated observations using other dummy device modules and servers.

\begin{figure}[H]
\centering
\includegraphics[width=0.8\textwidth, angle=0]{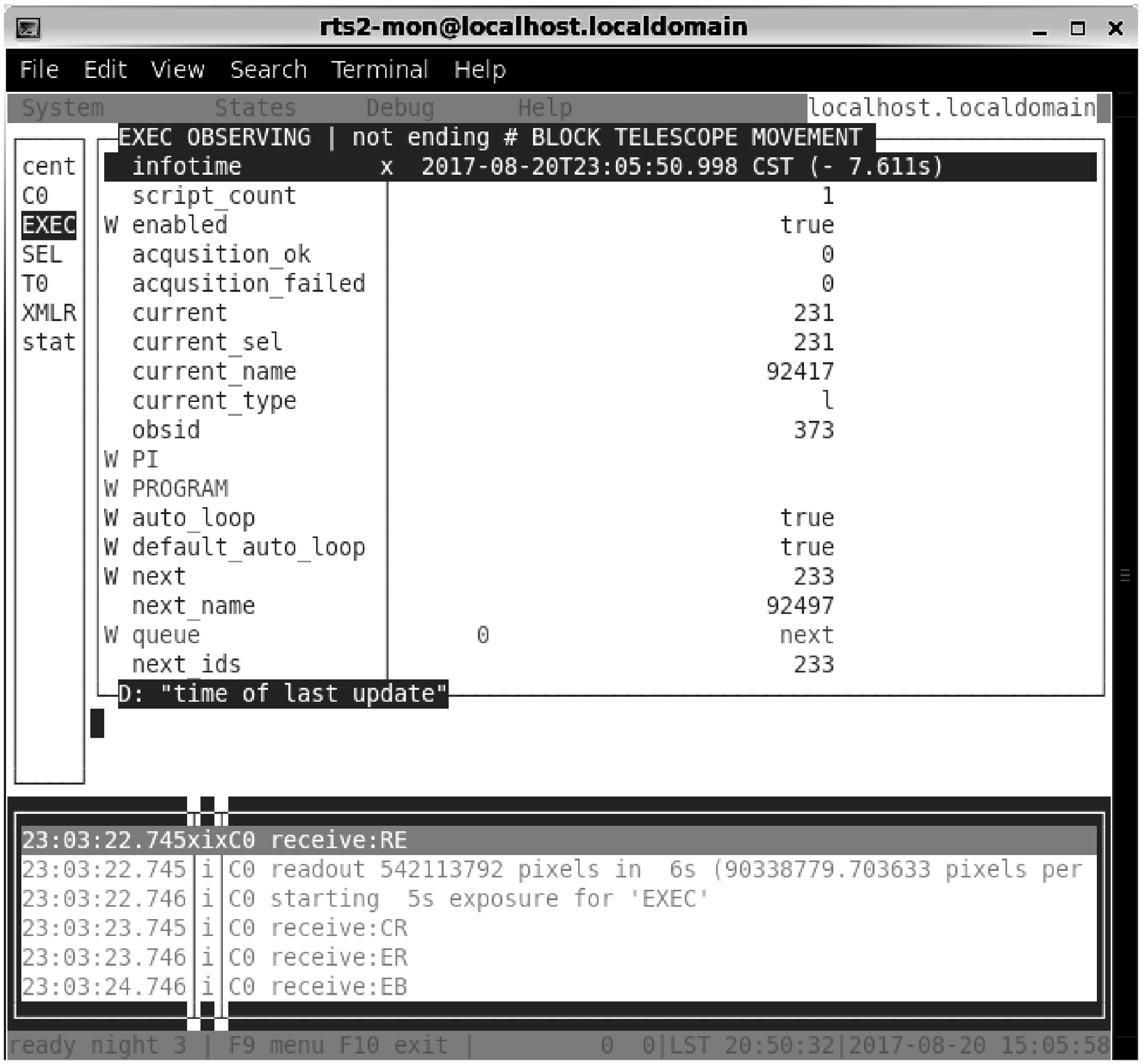}
\caption{Observation using new camera-control software with RTS2 environment}
\label{Fig12}
\end{figure}

We performed multiple observation tests continuously during a LAMOST maintenance session,  using a calibration lamp instead of real observation targets. A three-loop observation was performed for each target, and a custom script was run for each observation to yield exposures of 3 times and 5 seconds. Figure 11 shows the procedure of one of these tests. In the observation experiment, we created a target table in a PostgreSQL database and ran the RTS2 automatic mode. Then, RTS2 SEL selected a suitable target from the database and sent the target identifier to EXEC. EXEC then drove the telescope (T0 in Figure 11), which was an RTS2 dummy device module in this case. Our new camera-control system (C0 in Figure 11) worked with EXEC to complete the observation task. 

Figure 12 is a screenshot of the GUI of the new system for a given observation. This GUI can provide more detailed information on the statuses of the 32 EEV CCD cameras during operation than the existing system.

\begin{figure}[htbp]
\centering
\includegraphics[width=\textwidth, angle=0]{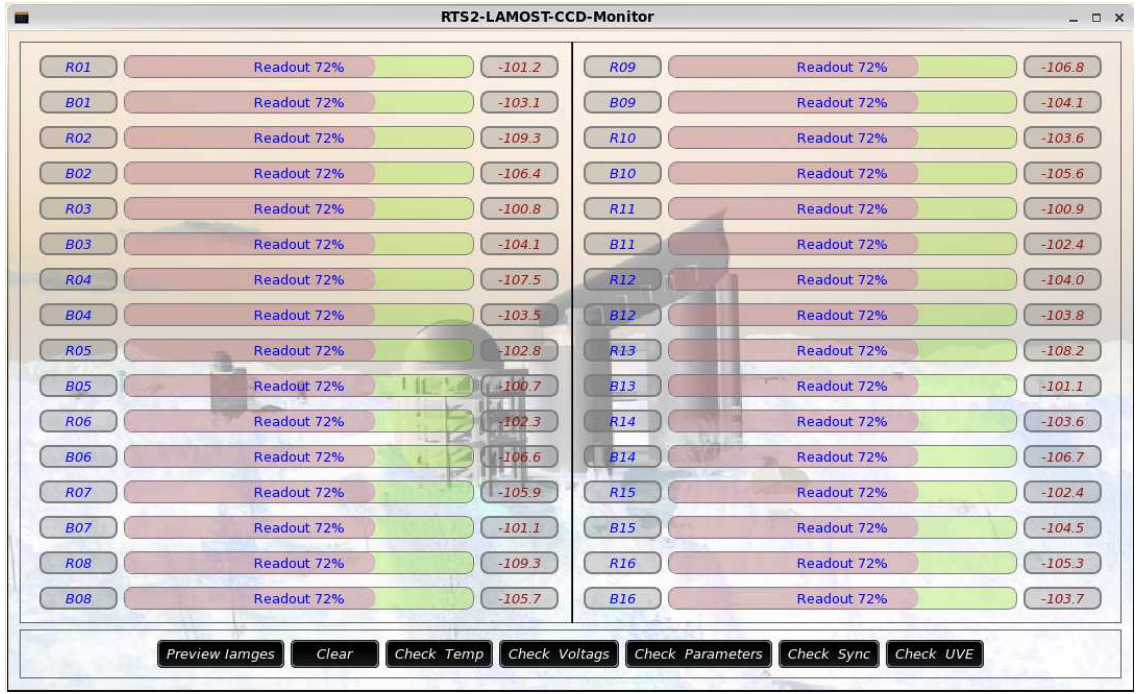}
\caption{GUI of new camera-control software during observation}
\label{Fig13}
\end{figure}

Figure 13 shows the results of a selected observation. Note that, after one observation, we obtain 32 target Flexible Image Transport System (FITS) files (Arc FITS files, to be specific).

\begin{figure}[!htbp]
\centering
\includegraphics[width=\textwidth, angle=0]{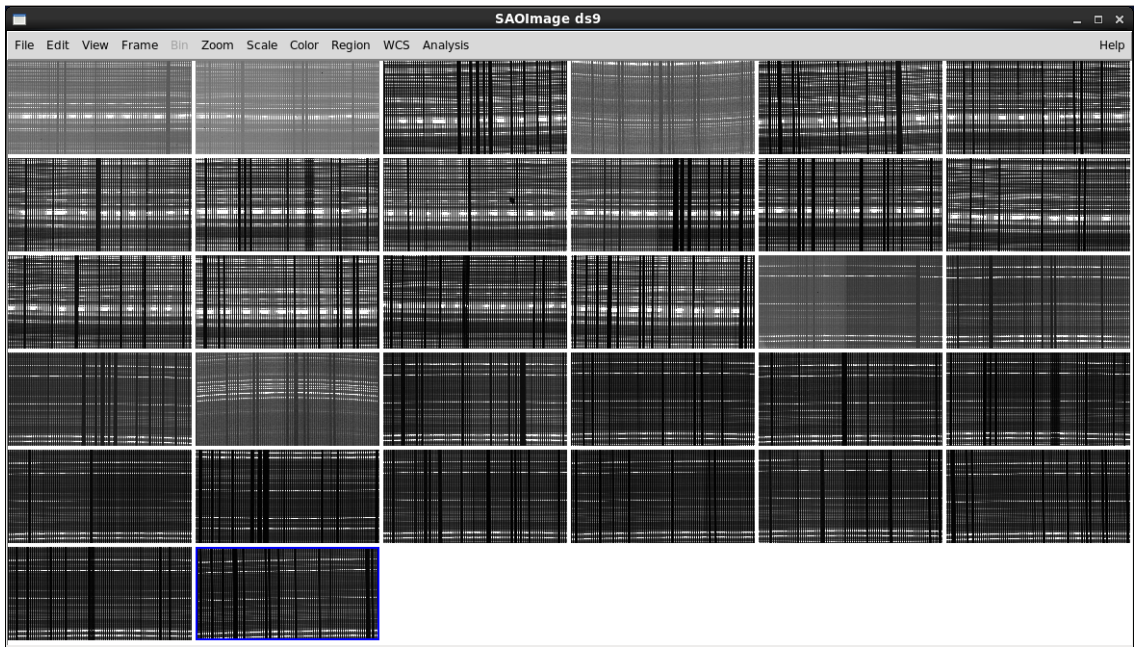}
\caption{Results provided by new system for one observation}
\label{Fig14}
\end{figure}

The results of the observation experiment indicate that the new RTS2-based camera-control software runs automatically and stably, and satisfies the LAMOST observation requirements.

\section{Discussion}
\label{sect:discussion}
\subsection{Virtual camera or 32 camera modules added to RTS2 framework}
To create the customized RTS2 device module for LAMOST, two design schemes were considered, as discussed below.
\subsubsection{First scheme}
According to this scheme, we would create 32 of our own camera modules, all of which would inherit from the RTS2::Camera class and overwrite certain interfaces (virtual functions). The number of customized devices would simply be increased. This process appears simple; however, there are many hidden disadvantages and risks. RTS2 was originally designed for small telescopes, and the intermodule communication mechanism uses a TCP-based network structure topology (not a star structure topology). When we register new modules in RTS2, the number of intermodule connections increases geometrically. An excessive number of connections not only reduces the stability of the RTS2 system, but also occupies additional bandwidth (many misuse messages are transferred between the new camera modules).
\subsubsection{Second scheme}
According to this scheme, a virtual layer that bridges RTS2 and the 32 real cameras is added. In this scheme, we create a customized camera and register it in the RTS2 framework like a simple camera. Then, we add a virtual layer (proxy) responsible for command distribution, coordination of the 32 real cameras, and status collection. This virtual layer simply provides an aggregate message to the RTS2 customized camera as feedback. By adding the virtual layer, we simplified the development process and reduced the coupling between the RTS2 and 32 real cameras. Another benefit also exists, i.e., we can reconstruct and upgrade the existing LAMOST CCD control software to reduce the workload and avoid development duplication.
Based on the above (and the previous sections), it is apparent that we chose the second scheme for implementation of the LAMOST camera-control system.

\subsection{Use of TCP instead of UDP}
In the existing camera-control software, a UDP broadcast mechanism is used, with the expectation that commands can arrive at each camera simultaneously. In order to improve the UDP reliability, many auxiliary codes have been added. However, packet loss and out-of-order problems always occur (~\cite{dong2009}). This problem is particularly serious in real observation scenarios, especially for cases involving large information throughput or a heavy network load. In addition, these auxiliary codes render the software obscure and increase the maintenance difficulty.

On the other hand, TCP provides reliable, ordered, and error-checked delivery of a stream of octets. This approach can overcome the shortcomings of UDP unreliability in the LAMOST high-speed local area network. From a technical perspective, TCP has lower efficiency than UDP. However, experiments have shown that careful TCP design can yield efficiency close to that for UDP. This is completely in line with the LAMOST engineering requirements (see Section 4.1 and Section 4.2). Further, use of TCP as the transport protocol can render the software both simple and stable. Moreover, use of TCP simplifies integration of the RTS2 framework into the new LAMOST camera-control system.

In summary, in the new RTS2-based LAMOST camera-control system, the UDP-based communication mechanism is replaced with a TCP transport protocol.

\subsection{Fault tolerance and exception handling mechanism}
 As an engineering project, the fault tolerance and exception handling mechanism is very important, it ensures the stable operation of the entire software system and provides human-computer interaction mechanism if necessary.The fault tolerance and exception handling mechanism of the new RTS2-based LAMOST camera-control system is in the fllowing aspects:
 
Firstly, as a device module of RTS2 framework, the new system inherits the fault tolerance of RTS2 directly. All RTS2 programs are designed as fault tolerant. The failure of one device does not affect other devices executing daemons(~\cite{peter2006}). It is possible to remove and add our virtual camera module during observation without affecting whole observation.

Secondly, in CCD-Master, status processor allocates a timer for each CCD-Slave to track the key operation steps regularly. For example, in reading process, each CCD-Slave transfers a message of readout progress (generated by UCAM) every second. The timers monitor these messages and create a warning information, if one CCD-Slave do not update the message within the specified time. The warning information will be broadcast to other RTS2 modules via RTS2's message mechanism( logStream() function). And it also can drive GUI to pop-up message box to remind the observer.

Thirdly, follow the RTS2 connections paradigms, the connection between CCD-Slave and CCD-Master is allows reconnection. When a single CCD camera encounters serious hardware problems, engineer can close this CCD-Slave and remove the connection. And after the problem is solved, CCD-Slave can be started and the connection will be reconnected is re-established automatically. This process does not affect other normal CCD operation.

The above considerations of fault tolerance and exception handling measures provide the necessary guarantees for the stable operation of the new camera-control system.

\subsection{RTS2 extension to entire LAMOST OCS}
Virtualization of complex devices is a very innovative concept for upgrading the control software of a large-scale astronomical telescope. This approach can add the features of the new software framework while retaining the interfaces of the existing software with the minimum possible changes. Hence, the software design is simplified and bugs caused by adding too much new code are reduced, with development being accelerated.

In order to improve the LAMOST automatic observation capability, we introduced the RTS2 framework to the LAMOST camera-control software. The 32 CCD cameras of LAMOST were mapped into one RTS2 virtual camera module. The results of the validation experiments conducted in this study prove the automatic observation capability of the new software.

The LAMOST OCS is an extremely large and complex software system, with the camera-control system being only one subsystem. To realize automatic observation capability for LAMOST, we must reconstruct other OCS subsystems, i.e., the tracking controller, fiber-positioning controller, etc. The architecture diagram for extension of RTS2 to the entire LAMOST OCS is shown in Figure 14.

\begin{figure}[!htbp]
\centering
\includegraphics[width=\textwidth, angle=0]{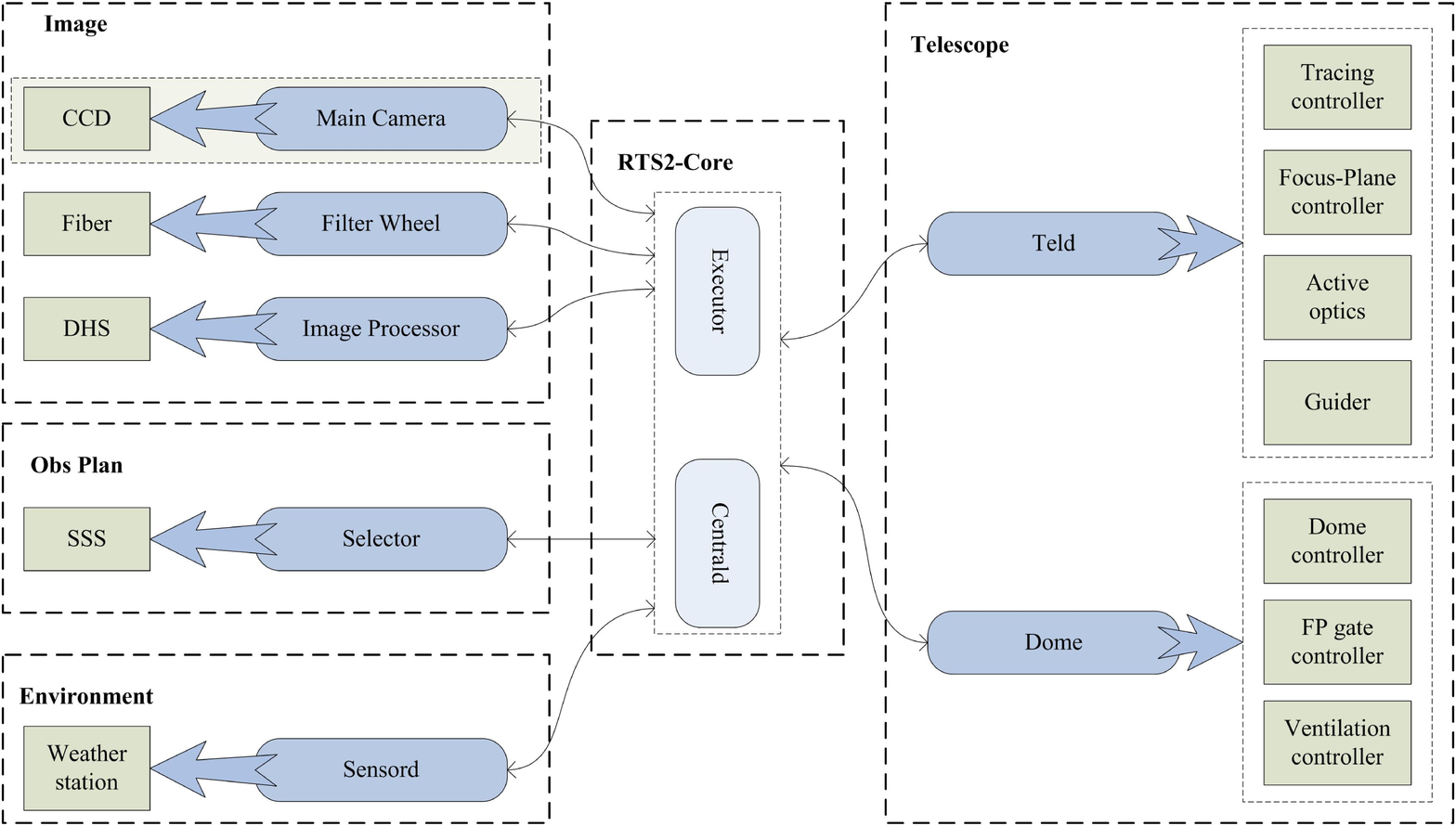}
\caption{Architecture diagram for extension of RTS2 to the entire LAMOST OCS}
\label{Fig15}
\end{figure}

Using the device virtualization approach, we will create additional virtual device modules to communicate with other existing OCS subsystems. Finally, by extending the RTS2 framework to the entire OCS, we will realize the ultimate goal of using RTS2 to control LAMOST.

Currently, modular design and layered implementation is the general trend for software development, and a large number of excellent new software frameworks are being used to implement this concept. To exploit the advanced features provided by these frameworks, compatibility issues between them and existing software must be addressed. The concept of existing device or software virtualization is an important method for resolving this problem. We believe that this concept is the guiding principle for introduction of the RTS2 framework to LAMOST and enhancing the automatic observation capability. In addition, this concept and its implementation will provide a reference for other large astronomical telescope software upgrades. .

\section{Conclusions}
\label{sect:conclusion}
In this study, we designed and implemented a new camera-control system based on RTS2 for LAMOST. The results of a series of experiments and observation tests prove the automation and stability of the new system, and indicate that the new system satisfies the observation requirements of LAMOST. In developing the new system, we reconstructed the existing software and implemented the concept of a virtual layer. These technologies greatly simplified the software development. Further, the approach described in this study will also provide a reference for software upgrades of other large-aperture astronomical telescopes, with the aim of realizing automation.

The camera-control system is only one part of the LAMOST OCS. In the future, other existing software packages (the mount, guider, online data handler, etc.) will be reconstructed for adaptation to the RTS2 framework. By retaining the concept of device virtualization, we are certain this task will be accomplished.

\begin{acknowledgements}
This study is supported by the National Key Research and Development Program of China (Grant No. 2016YFE0100300), the Joint Research Fund in Astronomy (Grant Nos. U1531132  U1631129 U1231205) under cooperative agreement between the National Natural Science Foundation of China (NSFC) and the Chinese Academy of Sciences (CAS), the National Natural Science Foundation of China (Grant Nos. 11603044 11703044 11503042 11403009 11463003).  The Guo Shou Jing Telescope ( the Large Sky Area Multi-Object Fiber Spectroscopic Telescope, LAMOST) is a National Major Scientific Project built by the Chinese Academy of Sciences. Funding for the project has been provided by the National Development and Reform Commission. LAMOST is operated and managed by the National Astronomical Observatories, Chinese Academy of Sciences. We also thank the reviewers for suggestions that improved the paper. The authors also gratefully acknowledge the helpful comments and suggestions of the reviewers.
\end{acknowledgements}


\label{lastpage}

\end{document}